\documentclass[hidelinks]{llncs}

\usepackage{amssymb}
\usepackage{mathtools}
\usepackage{dashbox}
\usepackage{algorithm}
\usepackage{algpseudocode}
\usepackage{setspace}
\usepackage{multirow}
\usepackage[binary-units=true]{siunitx}
\usepackage{subcaption}
\usepackage[nolist]{acronym}
\usepackage[htt]{hyphenat}
\usepackage{longtable}
\usepackage{booktabs}

\algnewcommand\algorithmicswitch{\textbf{switch}}
\algnewcommand\algorithmiccase{\textbf{case}}
\algnewcommand\algorithmicassert{\texttt{assert}}
\algnewcommand\Assert[1]{\State \algorithmicassert(#1)}%
\algdef{SE}[SWITCH]{Switch}{EndSwitch}[1]{\algorithmicswitch\ #1\ \algorithmicdo}{\algorithmicend\ \algorithmicswitch}%
\algdef{SE}[CASE]{Case}{EndCase}[1]{\algorithmiccase\ #1}{\algorithmicend\ \algorithmiccase}%
\algtext*{EndSwitch}%
\algtext*{EndCase}%

\usepackage{soul}
\newcommand{\ignore}[1]{}

\usepackage{tikz}
\usetikzlibrary{calc, positioning, arrows}
\tikzstyle{every node}=[node distance=2.25cm]

\usepackage{listings}

\newcolumntype{?}{!{\vrule width 1.25pt}}

\begin{document}

\title{Identifying Minimal Changes in the Zone Abstract Domain}

\author{%
  Kenny Ballou \orcidID{0000-0002-6032-474X} \and
  Elena Sherman \orcidID{0000-0003-4522-9725}
}
\institute{%
  Boise State University
  \email{kennyballou@u.boisestate.edu,elenasherman@boisestate.edu}
}
\authorrunning{Ballou, et al.}

\begin{acronym}
  \acro{AI}{abstract interpretation}
  \acro{CFG}{control flow graph}
  \acro{DFA}{data-flow analysis}
  \acro{DFS}{depth-first search}
  \acro{DBM}{difference bounded matrix}
  \acrodefplural{DBM}{difference bounded matrices}
  \acro{JVM}{Java virtual machine}
  \acro{LII}{linear integer inequalities}
  \acro{LIA}{linear integer arithmetic}
  \acro{TVPI}{two variables per inequality}
  \acro{SMT}{satisfiability modulo theories}
  \acro{StInG}{Stanford Invariant Generator}
  \acro{FS}{full state}
  \acro{FG}{Full Graph}
  \acro{CC}{Connected Components}
  \acro{NN}{Node Neighbors}
  \acro{MN}{Minimal Neighbors}
  \acro{NP}{Non-Polynomial}
\end{acronym}



\maketitle{}
\begin{abstract}
  Verification techniques express program states as logical formulas over
  program variables.  For example, symbolic execution and abstract
  interpretation encode program states as a set of linear integer inequalities.
  However, for real-world programs these formulas tend to become large, which
  affects scalability of analyses.  To address this problem, researchers
  developed complementary approaches which either remove redundant inequalities
  or extract a subset of inequalities sufficient for specific reasoning,
  \textit{i.e.,} formula slicing.  For arbitrary linear integer inequalities,
  such reduction approaches either have high complexities or over-approximate.
  However, efficiency and precision of these approaches can be improved for a
  restricted type of logical formulas used in relational numerical abstract
  domains.  While previous work investigated custom efficient redundant
  inequality elimination for Zones states, our work examines custom semantic
  slicing algorithms that identify a minimal set of changed inequalities in
  Zones states.

  The client application of the minimal changes in Zones is an empirical study
  on comparison between invariants computed by data-flow analysis using Zones,
  Intervals and Predicates numerical domains.  In particular, evaluations
  compare how our proposed algorithms affect the precision of comparing Zones
  vs. Intervals and Zones vs. Predicates abstract domains.  The results show
  our techniques reduce the number of variables by more than \(70\%\) and the
  number of linear inequalities by \(30\%\), comparing to those of full states.
  The approach refines the granularity of comparison between domains, reducing
  incomparable invariants between Zones and Predicates from \(52\%\) to
  \(4\%\), and increases equality of Intervals and Zones, invariants from
  \(27\%\) to \(71\%\).  Finally, the techniques improve the comparison
  efficiency by reducing total runtime for all subject comparisons for Zones
  and Predicates from over four minutes to a few seconds.

\end{abstract}



\section{Introduction}\label{sec:intro}%

Many verification techniques express a program state as a logical formula over
program variables.  For example, symbolic execution uses a logical formula to
describe a path constraint; in \acl{AI}, relational domains such as Zones or
Polyhedra use a set of \acl{LII} to describe program invariants.  While
expressive, this logical representation can become quite difficult to handle
efficiently.  For example, when symbolic execution traverses deep paths, or
when relational domains encode numerous program variables.  The increase in
formula size causes verification tasks to run out of memory or timeout.  Thus,
to improve scalability, verification techniques need to efficiently handle
these large logical formulas.

To overcome this predicament, researchers consider two complementary
approaches.  The first one focuses on eliminating the number of redundant
constraints using techniques such as Motzkin-Chernikova-Le
Verge~\cite{Chernikova:1965,Leverge:inria92} algorithm for \acl{LII}.  Previous
work, for example, used it to minimize path constraints in symbolic
execution~\cite{lloyd-2015-minim-size}.

The second approach focuses on identifying a minimal set of constraints
necessary to reason about a specific task, for example, identifying a set of
linear inequalities affected by state change.  Green~\cite{visser-2012-green}
performs the slicing operation on a path constraint formula to determine a set
of linear inequalities affected by a newly added constraint.  Identifying
relevant constraints reduces the query size sent to an \acs{SMT} solver to
check for satisfiability.

However, these minimization techniques assume a general format of linear
inequalities, which for a logical formula over \acl{LII} of restricted types,
may incur complexity cost or obtain a non-optimal solution.  In \acl{AI}, most
popular abstract relational domains such as Zones~\cite{mine-2001-new-numer} or
Octagons~\cite{mine-2006-octag-abstr-domain} restrict the type of linear
inequalities they encode.  Researches noted that Motzkin-Chernikova-Le Verge
algorithm has a high complexity and in some cases exponential
complexity~\cite{Yu:SAS19} when applied to eliminate redundant linear
inequalities.  Leveraging the efficient encoding for the Zones domain, Larsen
\textit{et~al.}~\cite{larsen-1997-effic-verif} developed a more efficient
algorithm which removes redundant constraints, with cubic complexity with
respect to the number of program variables.

The slicing technique proposed in Green uses syntax-based rules to compute
transitive dependencies of constraints.  While sound, this approach might
over-approximate the set of affected linear inequalities.  Applying a precise
``semantic-based'' slicing for a general linear inequality is a difficult
problem.  However, as this work shows for Zones domain, it reduces to quadratic
complexity.  In this work we propose several specialized algorithms for
computing a minimal changed set of linear inequalities for the Zones domain.
For efficient encodings and operations, relational numerical domains use
rewriting rules~\cite{Notzli:SAT19} to convert linear constraints into a
canonical form.  We also identify challenges such a canonical representation
causes in identifying minimal changes in an abstract state.

We evaluate our approach in the context of a \acf{DFA}
framework~\cite{kildall-1973-a-unified}, where \acl{AI} computes invariants
over program variables.  Researchers in areas such as program
verification~\cite{blanchet-2003-static-analyz,zhu-2018-a-data} or program
optimization~\cite{abate-2021-an-exten,katz-1978-progr-optim} use the computed
invariants to accomplish their respective goals.

The goal is to improve the precision of comparing invariants of Zones against
ones of Interval and Predicate abstract domains by comparing only the part of
Zone state that changed.  The importance of empirically evaluating domains has
been suggested previously~\cite{mine-2004-weakl-relat} since domains differ in
their expressiveness and efficiency, and thus, finding an optimal domain is an
important problem.


Evaluating our techniques to study the difference between incomparable domains,
\textit{e.g.,} Zones and Predicate domains~\cite{graf-1997-const-of}, allows us
to determine its effect on decreasing the number of incomparable comparisons
results.  For example, Zones can compute more precise values for some variables
at the beginning of a method, but later on, Predicates domain computes more
precise values for another set of variables.  If analyses are compared using
the entire state of each, then results would be incomparable for the later
part.  However, using our approach, the comparison would indicate that the
Predicates domain computes more precise invariants in the latter part of the
program.

Our main contributions for this work are:

\begin{itemize}
\item{A problem definition and a collection of efficient algorithms to identify
    minimal changes for the Zones abstract domain.}
\item{A demonstration of the effectiveness of our techniques at increasing
    precision when comparing Zones to comparable and incomparable domains.
    Similarly, demonstration that our techniques improve efficiency of domain
    comparison.}
\end{itemize}

\ignore{
The rest of the paper is organized as follows.  In Section~\ref{sec:backgrd},
we provide the background, context, and motivation for our work.  In
Section~\ref{sec:approach}, we define the problem and describe core
minimization algorithms.  In Section~\ref{sec:exper}, we explain our
experimental setup and evaluation, and in Section~\ref{sec:results}, we examine
the results of our experiments.  We connect this work with previous research in
Section~\ref{sec:related}.  Finally, we conclude and discuss future work in
Section~\ref{sec:concl}.
}



\section{Background and Motivation}\label{sec:backgrd}%

\begin{figure}[t]
  \centering
  \begin{subfigure}[b]{0.38\textwidth}
    \begin{spacing}{0.8}
      \centering
      \begin{lstlisting}[language=java,gobble=8]
        int example(int w, int y) {
          int x = 0;
          if (w <= x + 2) {
            if (y <= x) {
              assert y <= 0;
            }
          }
          return x;
        }
      \end{lstlisting}
    \end{spacing}
    \caption{Example program}\label{lst:example-progn}
  \end{subfigure}
  \begin{subfigure}[b]{0.3\textwidth}
    \centering
    \centering
    \begin{tikzpicture}[->,>=stealth',shorten >=1pt, auto, node distance=0.5cm and 0.5cm,
      semithick, font=\scriptsize]
        \node [draw, circle, inner sep=3pt] (0) at (0, 0) {$Z_0$};
        \node [draw, circle, inner sep=3pt] (x) at (0, 1.5) {$x$};
        \node [draw, circle, inner sep=3pt] (y) at (-1.5, 0) {$y$};
        \node [draw, circle, inner sep=3pt] (w) at (1.5, 0) {$w$};

        \draw[->] (x) edge [bend right, left] node{$0$} (0);
        \draw[->] (0) edge [bend right, right] node{$0$} (x);
        \draw[->] (w) edge [bend right, above right] node {$2$} (x);
        \draw[->,dashed] (y) edge [bend left, above left] node {$0$} (x);
    \end{tikzpicture}
    \caption{Zone state}\label{fig:zone-state-1}
  \end{subfigure}
  \begin{subfigure}[b]{0.3\textwidth}
    \centering
    \begin{tikzpicture}[->,>=stealth',shorten >=1pt, auto, node distance=0.5cm and 0.5cm,
      semithick, font=\scriptsize]
        \node [draw, circle, inner sep=3pt] (0) at (0, 0) {$Z_0$};
        \node [draw, circle, inner sep=3pt] (x) at (0, 1.5) {$x$};
        \node [draw, circle, inner sep=3pt] (y) at (-1.5, 0) {$y$};
        \node [draw, circle, inner sep=3pt] (w) at (1.5, 0) {$w$};

        \draw[->] (x) edge [bend right, left] node{$0$} (0);
        \draw[->] (0) edge [bend right, right] node{$0$} (x);
        \draw[->] (w) edge [bend right, above right] node {$2$} (x);
        \draw[->,dotted] (w) edge [above] node {$2$} (0);
        \draw[->,dashed] (y) edge [bend left, above left] node {$0$} (x);
        \draw[->,dotted] (y) edge [above] node {$0$} (0);
    \end{tikzpicture}
    \caption{Fully closed Zone state}%
    \label{fig:zone-state-closed}
  \end{subfigure}
  \caption{Example program and two equivalent Zones}%
  \label{fig:example-zone-state}
\end{figure}
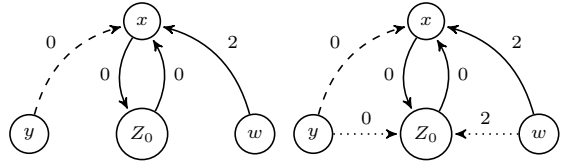

We illustrate problems with finding the minimal changed set of linear
inequalities for the Zones domain on a code example in
Figure~\ref{lst:example-progn} and focus on changes to the abstract state after
taking the true branch in statement 4, \textit{i.e.,} changes to the incoming
state of 4 to the outgoing state of the true branch of state 4.

To better conceptualize the idea of the minimal changed state, we first
consider abstract states computed by an analyzer over the
Intervals~\cite{cousot-1976-static} domain.  The incoming state is
\(x \mapsto [0, 0], w \mapsto (-\infty, 2], y \mapsto \top\), where the interval for
\(x\) comes from line 2; \(w\) comes from taking the true branch on line 3; and
\(y\), at this point, is unbounded.  After applying the transfer function of
4t, the analyzer updates the value of \(y\) to \(\left(-\infty, 0\right]\) without
affecting values of \(x\) and \(w\).  To identify changed variables, the
analyzer simply checks the difference in updated variable values between the
two states since changes to one variable do not induce changes in other
variables.

\subsection{Finding a Minimal Subset in Relational Domains}

For a relational numerical domain, the analyzer produces the formula
\(x = 0 \land w - x \le 2\) for the incoming state to line 4, the absence of the
\(y\) variable means it is a free variable.  Interpreting 4t, the analyzer
introduces: \(y - x \le 0\), resulting in a new outgoing state:
\(x = 0 \land w - x \le 2 \land y - x \le 0\).  Here, the minimal subset of
inequalities contains $x = 0$ and $y - x \le 0$.  Indeed, only these two
inequalities are sufficient to reason about the changed part of the state.

Such a minimal subset is not easily identifiable.  Green's slicing technique
over-approximates it by including all inequalities through a syntax-based
transitive closure.  However, since $x$ has not changed and is only used in
$y - x \le 0$ to restrict $y$'s values, \emph{no changes occurred with respect to
  $w$}, and hence, the $w - x \le 2$ inequality remains the same.  In our work,
we show algorithms for reasoning about such ``semantic-based'' slicing for the
Zones domain.  Before we provide an overview of our approach, we present a
brief background on the Zones domain.

\subsection{Zones Domain}

The Zones abstract domain expresses specific relations between program
variables.  Zones limits its relations to unitary difference inequalities such
as \(x - y \le b\), and interval inequalities such as \(x \le b\) and
\(x \ge b\), where \(b \in \mathbb{Z}\).  Equality relations are rewritten into a pair of
two inequalities.  For example, an \(x = y + b\) relation results in
\(x - y \le b\) and \(y - x \le -b\).  To encode an interval value such as
\(x = [1, 2]\), Zones use the following inequalities \(x \ge 1\) and
\(x \le 2\).  To encode interval valued variables as unitary difference
constraints, Zones introduces a special ``zero'' variable, denoted here as
\(Z_0\), and rewriting rules.  These rules change the inequalities for the
interval \([1, 2]\) into \(x - Z_0 \le 2\) and \(Z_0 - x \le -1\) where the value
of \(Z_0\) always equals \(0\).  In this way, the Zones domain represents all
inequality constraints in a uniform \(x - y \le b\) template.

Inequalities in the Zones domain have an isomorphic representation as a
weighted, directed graph, which is efficiently encoded as a 2D
matrix~\cite{dill-1990-timin,mine-2001-new-numer}.  In this graph, variables
are nodes, the source and the sink of an edge identify variables in the first
and the second positions of the difference template, respectively, and the
weight is the coefficient \(b\).

To illustrate this representation, consider the graph in
Figure~\ref{fig:zone-state-1}, which encodes the constraints from the running
example.  The solid lines denote the inequalities of the incoming state to
statement 4; and the dashed line is the additional inequality after 4t.  Here,
the edge from \(x\) to \(Z_0\) is for \(x - Z_0 \le 0\) and the edge from
\(Z_0\) to \(x\) is for \(Z_0 - x \le 0\) of the rewritten \(x = 0\) term.  The
edge from \(w\) to \(x\) represents \(w - x \le 2\).  Similarly, the dashed edge
from \(y\) to \(x\) represents the additional \(y - x \le 0\) term.

\subsection{Finding Affected Inequalities}

In graph terms, the problem reduces to finding affected edges when an edge is
changed in the state graph.  Naively, all edges reachable from the changed
variables are affected, \textit{i.e.,} a connected component (with undirected
edges).


However, we should exclude nodes connected only through $Z_0$ as a part of the
connected component since it might introduce \emph{spurious} dependencies.
Consider, the state also has an edge from a variable \(w\) to \(Z_0\) with
weight \(2\), which encodes inequality \(w \le 2\) represented as
\(w - Z_0 \le 2\) in the canonical unitary difference form.  If $Z_0$ is treated
as a regular variable, then the technique would include \(w - Z_0 \le 2\) in the
set of affected inequalities.  Our approach considers \(Z_0\) a singularity and
stops at \(Z_0\) when it transitively computes predecessors \(R^-\) and
successors \(R^+\) of a node, for example, to identify connected components.

To improve over the connected component approach, our algorithms reason about
the directions of the edges in the graph.  The outgoing edges from one node to
another means the second one restricts the first.  In our example, $w$ has an
outgoing edge to $x$, meaning $x$ restricts $w$.  The only changes that can be
propagated to $w$ from an additional inequality, if $x$'s value would be
further restricted, manifests as a new or updated outgoing edge.  However,
after adding \(y - x \le 0\) inequality (the dashed edge), the only new outgoing
transitions are associated with $y$.  Thus, in order to determine the affected
inequalities we transitively compute the predecessors and successors of \(y\),
\textit{i.e.,} \(R^-(y) = \{y\}\) and \(R^+(y) = \{x, y, Z_0\}\).  The union of
the two sets results in the subgraph of dependent variables to \(y\).  In the
next section we present formal algorithms and prove their correctness.


\subsection{Dealing with Spurious Inequalities}

Relational numerical domains possess a powerful ability to infer new relations
between variables.  In the Zones domain, it happens through computing
transitive closures for each variable.  For example, since $w$ can reach $Z_0$,
one can establish that $w \le 2$.  Later, when the \(x\) variable gets
reassigned, $w \le 2$ remains in place.  Since inequalities describing a bounded
region are not unique, Zones require a canonical representation to enable
operations such as equality comparisons.  In order to determine whether two
Zones are equal, as needed in \acs{DFA}'s fixed-point algorithms, these states
should be \emph{fully closed}.  That is, all inferred, transitive constraints
should be explicitly stated.  Figure~\ref{fig:zone-state-closed} shows the
fully closed version of the same graph as in Figure~\ref{fig:zone-state-1},
where dotted lines depict inferred edges.

Applying the previously described technique identifies three inequalities
\(y \le x\), \(x = 0\) and \(y \le 0\).  Clearly, the first two constraints do not
add any additional information to the third.  In fact, after closing, the newly
added edge $(y, x)$ becomes what we call a \emph{spurious} edge since it
falsely implies that $y$ and $x$ make up a relational constraint, while in
fact, since $x$ is a constant, this is not the case.

Therefore, before finding affected inequalities, we identify such spurious
edges and remove them from the graph.  The algorithm checks for a given
connected component containing changed variables whether an edge between two
nodes connected to $Z_0$ carries additional information, otherwise it removes
the direct edge.  In our example, the connected component containing
potentially changed variables are all nodes with an edge to $Z_0$.

In our example, the algorithm removes edges $(y, x)$ and $(w, x)$.  Now, node
$x$ is no longer reachable from $y$, making \(R^+(y) = \{y, Z_0\}\).  For a
fully closed state the algorithm determines a single affected inequality:
\(y \le 0\).  Thus, removing spurious constraints helps with identifying smaller,
truly connected components.

\section{Finding the Minimal Changed Set of Inequalities}\label{sec:approach}%

In this section, we first formally define the problem of finding a minimal
changed set of inequalities from Zones.  We continue with a series of
algorithms starting with spurious constraint elimination, followed by different
minimization approaches for arbitrary and fully closed Zones.


\subsection{Problem Definition}

Let \(\mathcal{N}_1\) and \(\mathcal{N}_2\) be sets of inequalities in the initial and updated
states, respectively.  An inequality \(n\) of a Zone state can be uniquely
identified by its first \(n_1\) and the second \(n_2\) variables in the unitary
difference formula template.  Let \(u\) be the node representing \(n_1\) and
similarly \(w = n_2\), then the corresponding edge \(n\) in the state graph has
\(u\) as its source and \(w\) as its target nodes.

For a given variable \(v\), we define its dependent inequalities in \(\mathcal{N}_1\) as
\[S_1 = \{ n \in \mathcal{N}_1 ~|~ n_1 \in R_{\mathcal{N}_1}^-(v) \lor n_2 \in R_{\mathcal{N}_1}^+(v) \}\] where
\(R_{\mathcal{N}_1}^-(v) = P(v) \cup \{v\}\), and similarly,
\(R_{\mathcal{N}_1}^+(v) = S(v) \cup \{v\}\).  \(S_2\) is dually constructed for \(\mathcal{N}_2\).


Let \(dv\) be a nonempty set of updated variables that change \(\mathcal{N}_1\) to
\(\mathcal{N}_2\).  Then, the problem of finding the minimal changed set of inequalities
is equivalent to finding the smallest \(S \subseteq \mathcal{N}_2\) such that
\[S \cap \{ n \in \mathcal{N}_2 ~|~ \forall v \in dv: n_1 \in R_{\mathcal{N}_2}^-(v) \lor n_2 \in R_{\mathcal{N}_2}^+(v)\} \neq \emptyset
  \text{ and } S_2\setminus_{id} S \Leftrightarrow S_1\setminus_{id} S\]

Here, the first term ensures that the set \(S\) includes only affected
inequalities.  The second term ensures that the set of remaining inequalities
after removing updated ones should be logically equivalent between the initial
and updated states.  The equality in the set minus operation is determined by
inequality IDs, \textit{i.e.,} \(n_1\) and \(n_2\).  That is, if two
inequalities have the same order of variables, then they are considered
equivalent for set operations.

\subsection{Minimization Algorithms}

A straightforward solution to identify such an \(S\) would be to compare
\(\mathcal{N}_1\) and \(\mathcal{N}_2\) directly, which would require an exhaustive search.
Instead, our approaches only take \(\mathcal{N}_2\) as a graph \(Z\), a set of updated
variables \(dv\), and a set of updated edges \(de\).  A \acs{DFA} framework can
directly provide the sets \(dv\) and \(de\) when it invokes a transfer
function.  Using these three input values and a choice of minimization
\(Method\), the pseudocode in Algorithm~\ref{alg:minimal-subset} computes the
smallest set of changed inequalities which it returns as a graph \(G\).



\begin{figure}[t]
  \centering
  \begin{algorithm}[H]
    \scriptsize
    \begin{spacing}{1.0}
      \begin{algorithmic}[1]
        \Function{MinChangedSet}{$Z$, $dv$, $de$, $Method$}
          \State{$G \gets $ \Call{RemoveSpurious}{$Z$}}\label{alg:minimal-subset-woz}
          \Switch{$Method$}\label{alg:minimal-subset-cond}
          \Case{$CC$}
          \State{$G \gets $ \Call{ConnectedComponents}{$G$, $dv$}}\label{alg:minimal-subset-cc}
          \EndCase{}
          \Case{$NN$}
          \State{$G \gets $ \Call{NodeNeighbors}{$G$, $dv$}}\label{alg:minimal-subset-nn}
          \EndCase{}
          \Case{$MN$}
          \State{$G \gets $ \Call{MinimalNeighbors}{$G$, $de$}}\label{alg:minimal-subset-mn}
          \EndCase{}
          \EndSwitch{}
          \State{\Return $G$}
        \EndFunction{}
      \end{algorithmic}
    \end{spacing}
  \caption{Minimal Changed Set}\label{alg:minimal-subset}
  \end{algorithm}
  \vspace{-20pt}
\end{figure}

On line~\ref{alg:minimal-subset-woz}, the algorithm invokes
\texttt{RemoveSpurious} on the updated state \(Z\). The purpose of this method
is to remove spurious dependencies in \(Z\) inferred through \(Z_0\), thus
creating smaller connected components in $G$.  Next, on
line~\ref{alg:minimal-subset-cond}, the algorithm switches on $Method$.  The
rest of the algorithm computes minimal changed sets based on the method
selected: \acs{CC} approach-- \texttt{ConnectedComponents}; \acs{NN} approach--
\texttt{NodeNeighbors}, and lastly \acs{MN}-- \texttt{MinimalNeighbors}. These
algorithms approximate $S$ differently and have different computational
complexity, which are varied based on the Zones representation.  $G$s of these
algorithms create a total order on the number of inequalities in the reduced
state $G_{CC} \preceq_{|E|} G_{NN} \preceq_{|E|} G_{MN}$.  The runtime complexity of
Algorithm~\ref{alg:minimal-subset} is \(O(n^2)\),\footnote{The average is
  usually less due to sparsity in graphs.} which is dominated by the quadratic
complexity of \texttt{RemoveSpurious}.

\subsubsection{Spurious Connections}\label{sec:approach-woz}

\begin{figure}[t]
  \centering
  \begin{algorithm}[H]
    \scriptsize
    \begin{spacing}{1.0}
      \begin{algorithmic}[1]
        \Function{RemoveSpurious}{$G$}\label{alg:woz-reduction-parameters}
          \State{\(C \gets \{\}\)}
          \For{\(s \in \) \Call{V}{\(G\){}}}
            \If{\((s, Z_0) \neq \top \lor (Z_0, s) \neq \top\)}
              \State{\(C \gets C \cup \{s\}\)}
            \EndIf{}
          \EndFor{}
          \For{\(s \in C\)}
            \For{\(t \in C\)}\label{alg:woz-reduction-iteration}
              \If{\(s \ne t \land (s, t) \ge (s, Z_0) + (Z_0, t)\)}\label{alg:woz-reduction-comparison}
                \State{\((s, t) \gets \top\)}\label{alg:woz-reduction-removal}
              \EndIf{}
            \EndFor{}
          \EndFor{}
          \State{\Return $G$}
        \EndFunction{}
      \end{algorithmic}
    \end{spacing}
    \caption{Removal of spurious dependences in G}\label{alg:woz-reduction}
  \end{algorithm}
  \vspace{-20pt}
\end{figure}

The goal of the spurious constraint removal step is to deal with inferences
through \(Z_0\).  Our approach for identifying spurious edges is a special case
of the reduction proposed by Larsen
\textit{et~al.}~\cite{larsen-1997-effic-verif}.  That is, an edge between two
nodes can be removed from a Zone if the weight between them is greater or equal
to any path between them.  Instead of applying this reduction to the entire
state, our algorithm considers only the path \textit{through} \(Z_0\). This
reduces the runtime complexity from \(O(n^3)\) to \(O(n^2)\).  We define a
spurious directed edge between \(s\) and \(t\) variables when
\((s, t) \ge (s, Z_0) + (Z_0, t)\).

Algorithm~\ref{alg:woz-reduction} details the steps for this spurious edge
removal.  The algorithm determines candidate pairs by selecting variables with
connections to or from \(Z_0\), line~\ref{alg:woz-reduction-parameters}.  Nodes
not connected to the zero node can be excluded because any edge is, by
definition, non-spurious.  It iterates over the candidate node pairs,
line~\ref{alg:woz-reduction-iteration} and for each pair, it checks the
spurious edge criterion on line~\ref{alg:woz-reduction-comparison}.  If the
criterion is satisfied, the edge is removed,
line~\ref{alg:woz-reduction-removal}.  The correctness of the algorithm comes
from the spurious edge criterion: it never removes an edge inferred by \(Z_0\)
that is not redundant.  Furthermore, spurious edges represent redundant
constraints, therefore, removing them does not cause precision loss.





\subsubsection{Connected Components}\label{sec:connected-components}

For an arbitrary Zone, we can safely over-approximate all affected inequalities
from \(dv\) by identifying a connected component containing \(dv\).  Note, that
changed variables are always in one component, since an update to a Zone
creates an edge between them.  Identifying \(dv\)'s connected component reduces
to discovering the undirected, reachable nodes of each \(v \in dv\) in the
spurious reduced Zone.  The \texttt{CC} algorithm is a modified \acl{DFS}
algorithm, with \(O(n^2)\) runtime complexity.  In the beginning, the algorithm
marks the special \(Z_0\) node as visited, thus, preventing discovery of new
paths through it, lest we undo the reduction of
Algorithm~\ref{alg:woz-reduction}.

The \acs{CC} algorithm is the same for arbitrary, non-closed Zones and fully
closed Zones.  The resulting sets of changed variables, however, may differ.
The set of inequalities of fully closed Zones are minimized but often more
connected.  However, more connections enables more spurious reductions,
therefore, leading to smaller connected components.

\subsubsection{Node Neighbors}\label{sec:approach-neighbors}

\begin{figure}[t]
  \centering
  \begin{algorithm}[H]
    \scriptsize
    \begin{spacing}{1.0}
      \begin{algorithmic}[1]
        \Function{NodeNeighbors}{$G$, $dv$}
          \State{variables $\gets \{\}$}
          \For{$v \in dv$}\label{alg:general-nn-iteration}
            \State{variables $\gets $ variables $ \cup $ \Call{ForwardReachable}{$G$, $v$}}
            \State{variables $\gets $ variables $ \cup $ \Call{BackwardReachable}{$G$, $v$}}
          \EndFor{}
          \State{\Return{} variables}
        \EndFunction{}
      \end{algorithmic}
    \end{spacing}
    \caption[General Node Neighbor Algorithm]{Algorithm for node neighbor
      selection for arbitrary Zones.}%
    \label{alg:general-node-neighbors}
  \end{algorithm}
  \vspace{-20pt}
\end{figure}

The \acf{NN} algorithm for an arbitrary Zone state is presented in
Algorithm~\ref{alg:general-node-neighbors}.  In essence, the algorithm searches
for the successor and predecessor of each changed variable,
line~\ref{alg:general-nn-iteration}.  \texttt{ForwardReachable} returns the set
of all reachable successor variables for each changed variable, \(v \in dv\),
using a typical \acl{DFS}.  \texttt{BackwardReachable} is similarly defined for
reachable predecessor variables.  In both cases, the zero variable receives
special consideration.  That is, we specially treat the zero variable as a sink
with no outgoing edges during traversal.

The complexity of \acs{NN} is \(O(4n^2)\) because there are at most \(2\)
changed variables from the \acs{DFA} framework, and we do \acs{DFS} twice per
variable.


When given a fully closed Zone, the \acs{NN} complexity is reduced.  The form
of a fully closed Zones makes explicit all transitively related variables.
That is, the set of successors and predecessors of a variable in a fully closed
Zone is equivalent to the local neighborhood of the variable,
\(R_G^-(v) \cup R_G^+(v) = N_G^{\pm}(v)\).  Therefore, \acs{NN} for the fully closed
Zones simply returns the inequalities incident to the neighbor set of the
requested variable.

The complexity and accuracy of finding a minimized state can be reduced for
Zones in fully closed canonical form, where all dependencies are explicit.
Thus, to identify affected inequalities by $dv$, we need to find incoming and
outgoing edges of $dv$ since they are potentially affected by those updates.
The \acs{NN} algorithm takes $dv$ and a reduced state graph $G$, and retrieves
all incoming and outgoing edges of $dv$, then uses them to identify its
neighbors.  The local subgraphs for each \(v \in dv\) represent the identified
changed inequalities.  The runtime complexity of \acs{NN} is linear $O(n)$, since it
only considers each associated edge of $dv$.

The correctness of \acs{NN} relies on the dependency information encoded in the
successors and predecessors set.  If a variable \(u\) is not in the
\(R_G^-(v) \cup R_G^+(v)\) set for \(v\), then \(v\) does not depend on or relate
to \(u\).  Furthermore, since \(u\) is not in this dependency set, it has not
changed from the previous state.  Therefore, the variable \(u\) can be removed
from the dependent inequality set returned by \acs{NN}.

\subsubsection{Minimal Neighbors}\label{sec:approach-min-neighbors}

\begin{figure}[t]
  \centering
  \begin{algorithm}[H]
    \scriptsize
    \begin{spacing}{1.0}
      \begin{algorithmic}[1]
        \Function{MinNeighbors}{$G$, $de$}
          \State{variables $\gets \{\}$}
          \For{$(s, t) \in de$}\label{alg:min-neighbors-iteration}
            \If{$s = Z_0$}
              \State{variables $\gets \text{variables} \cup \{t\}$}\label{alg:min-neighbors-t}
            \ElsIf{$t = Z_0$}
              \State{variables $\gets \text{variables} \cup \{s\}$}\label{alg:min-neighbors-s}
            \ElsIf{$s \neq Z_0 \land t \neq Z_0$}
              \State{variables $\gets \text{variables} \cup \{s\}$}\label{alg:min-neighbors-st}
            \EndIf{}
          \EndFor{}
          \State{\Return $NodeNeighbors(G, variables)$}%
          \label{alg:min-neighbors-return}
        \EndFunction{}
      \end{algorithmic}
    \end{spacing}
    \caption{Minimize Changed Variables Algorithm}\label{alg:min-neighbors}
  \end{algorithm}
  \vspace{-20pt}
\end{figure}

The previous \acs{CC} and \acs{NN} minimization algorithms assume that all
updated variables, \(dv\), modify inequalities within a Zone state, however,
that may not always be the case.  An updated variable might not induce changes
to the state.  The \acf{MN} technique improves upon this over-approximation by
considering the set of updated edges \(de\) in a Zone state.  \acs{DFA}
framework can provide this information when processing a statement,
\textit{e.g.,} an assignment or a conditional statement.  Notice, that sources
and targets of an edge in \(de\) are always in \(dv\), but additional
computation is required to identify \(de\).

Algorithm~\ref{alg:min-neighbors} shows the pseudocode for identifying the
changed variables among updated edges.  Specifically, the algorithm takes as
input $G$ produced by \texttt{RemoveSpurious} and a set of updated edges,
\(de\).  It starts by iterating over the set of updated edges,
line~\ref{alg:min-neighbors-iteration}.  For each edge, the algorithm checks
whether the source or target is the zero node, \(Z_0\).  If so, then the other
node from the edge pair is added to the \texttt{variables} set,
lines~\ref{alg:min-neighbors-t} and~\ref{alg:min-neighbors-s}.  This case
handles when updates are on intervals because we must always include the non
$Z_0$ variable in the filtered set of changed variables.

If neither source nor target is \(Z_0\), then the source of the edge is added
to the \texttt{variables} set, line~\ref{alg:min-neighbors-st}, since this
corresponds to the target variable restricting the source node, while the
former remains unchanged for that edge.  Finally, on
line~\ref{alg:min-neighbors-return}, the algorithm invokes \acs{NN} procedure
on \(G\) and the computed changed set of variables, and returns the minimized
graph.  The runtime complexity of \acs{MN} is equivalent to \acs{NN}:
\(O(4n^2)\).  Similarly, if \(G\) is fully closed, the complexity is $O(n)$.


Below we provide a proof sketch that shows that for a given $G$ from
\texttt{RemoveSpurious} algorithm and an updated edge $(s,t)$, the variable of
its target, \(t\) does not change if the edges \((t, s)\) and \((Z_0, s)\) do
not exist.  That is, the target variable is never added to the variable set in
Algorithm~\ref{alg:min-neighbors}.

\begin{proof} Given \(G\) with an updated edge \((s, t)\) corresponding to the
  additional constraint \(s - t \le c\), for some \(c \in \mathbb{Z}\).  For the purpose of
  contradiction, let us assume \(t\) has changed as a result of the update from
  \(s - t \le c\).  This means either there exists a path from \(t\) to \(s\) or
  that there exists a path \((Z_0, s)\).  But existence of such paths violates
  our assumption that \(t, Z_0 \notin R^-_G(s)\).  Therefore, \(t\) remains
  unchanged by the addition of the edge \((s, t)\).
\end{proof}

\subsection{Widening and Merges}

A few situations require special treatment for state updates in \acs{DFA}.
First, widening and merge points in the \acs{CFG} of the analysis may induce
more changed variables.  Second, conditional transfers tend to modify more than
a single variable, \textit{e.g.,} two variables in three address form.
Therefore, to ensure accurate and minimal comparisons, our techniques and
comparisons must handle these situations.  However, this is easily accomplished
by each of our techniques since the parameter, \(dv\) for
Algorithm~\ref{alg:minimal-subset} is a set of changed variables.



\section{Experiments Methodology}\label{sec:exper}%

To determine if the proposed minimization algorithms are efficient and
effective, we evaluate them using subject programs within an existing \acs{DFA}
analyzer.  For each subject program, we compute invariants at each statement
for each abstract domain.  Over the corpus of methods, we compute \(4529\)
total invariants.  The invariants are stored as logical formulas in \acs{SMT}
format.  We run analysis on three domains: Zones, Intervals and Predicates, and
compare the first with the last two using queries to an SMT solver.  Since
previous research demonstrates advantages of using fully closed Zone
states~\cite{ballou-2022-increm-trans}, our experiments evaluate minimization
algorithms for that canonical representation.

To evaluate the efficacy of \acs{CC}, \acs{NN} and \acs{MN}, each after
computing spurious constraint reduction, we compute the reduction in the number
of variables and inequalities in the \acs{SMT} formula of Zone invariants over
the preceding techniques in $\preceq$, \textit{i.e.,} \acs{CC} vs. \acf{FS}, \acs{NN}
vs. \acs{CC}, and \acs{MN} vs. \acs{NN}.  That is, we compute the percentage of
change per program statement and then average them over all methods.  We use
percentage, and not absolute values since the number of variables changes from
statement to statement.  Similarly, we compute the average percentage change
for inequalities.  Since program branches compute possibly different sized sets
of variables and inequalities, we take the maximum number of variables and
inequalities between the two.


Using the invariants computed for Zones, Intervals, and Predicates, we entail
the invariants to compare the precision of Zones vs. Intervals and Zones
vs. Predicates for each minimization algorithm.  The results for Interval's
invariants are classified as less precise $\prec$ or equal $=$.  Predicates extend
these categories to include more precise $\succ$ or incomparable $\prec\succ$ to Zones.

\noindent{\bf{Subject Programs}} Subject programs consist of \(127\) Java
methods used in previous
research~\cite{badihi-2021-eqbench,sherman-2015-exploit-domain}.  Methods from
the \acs{DFA} benchmark suite~\cite{sherman-2015-exploit-domain} were extracted
from a wide range of open-source projects and have a high number of integer
operations.  The subject programs range from \(4\) to \(412\) Jimple
instructions, a three address intermediate representation.
The EQBench suite~\cite{badihi-2021-eqbench} consists of method pairs for
testing differential symbolic execution
tools~\cite{badihi-2020-ardiff,person-2008-differ-symbol-execut}.  We sampled
only original methods and excluded renamed equivalent methods.

\noindent{\bf{Experimental Setup}} We execute each of the analyses on a single
GNU/Linux machine, running Linux kernel version \texttt{5.15.89}, equipped with
an AMD Ryzen Threadripper 1950X 16-Core Processor and \SI{64}{\giga\byte} of
system memory.  We use an existing \acs{DFA} static analysis
tool~\cite{sherman-2015-exploit-domain} implemented in the Java programming
language.  The analysis framework uses Soot~\cite{web-soot-oss} version
\texttt{4.2.1}.  Similarly, we use Z3~\cite{moura-2008-z3}, version
\texttt{4.8.17} with Java bindings to compare \acs{SMT} expressions from the
abstract domain states.  Finally, we use Java version 11 to execute the
analyses, providing the following \acs{JVM} options: \texttt{-Xms4g},
\texttt{-Xmx32G}, \texttt{-XX:+UseG1GC}, \texttt{-XX:+UseStringDeduplication},
and \texttt{-XX:+UseNUMA}.


\noindent{\bf{Implementation}} We use the reduction from Larsen
\textit{et~al.}~\cite{larsen-1997-effic-verif} to create an equivalent, but
reduced invariant expression at each program point.  We combine the output
states via logical entailment to compare Zones to Intervals and to Predicates.
The set of variables in a minimized Zone state determines what variables are
extracted from the corresponding full states of Intervals and Predicates.
After entailment, we use Z3, using the \acf{LIA} logic, to decide model
behavior of each domain.  Using the GNU \texttt{time}~\cite{web-gnu-time}
command, version 1.9, we capture the walk-clock execution time of Z3.

\noindent{\bf{Evaluations}} Intervals and Zones perform widening operations
after two iterations over widening program points.  We do not preform narrowing
for either domain because narrowing is program specific.  The lack of narrowing
does not affect our results since we are evaluating techniques for identifying
minimal subsets of changes, not techniques for improving precision.

We use a generic disjoint Predicate domain, which does not affect generality of
the results.  The Predicate domain's elements are influenced by Collberg
\textit{et~al.'s}~\cite{collberg-2007-empir-study} study on Java programs and
numerical constants.  Consequently, the domain elements use several of the most
common integer constants found in Java programs.  The specific Predicate domain
used in this study consists of the following set of disjoint elements:
\(\{(-\infty, -5]\), \((-5, -2]\), \(-1\), \(0\), \(1\), \([2, 5)\), \([5, +\infty)\}\).



\section{Evaluation Results and Discussions}\label{sec:results}%

To empirically evaluate efficiency and effectiveness of the state minimization
algorithms, we answer the following research questions:


\begin{description}
\item[\textbf{RQ1}]{How well do the minimization algorithms reduce the size of
    a Zone state and improve runtime of domain comparisons?}
\item[\textbf{RQ2}]{How do the minimization algorithms affect categorization of
    domain comparison results?}
\end{description}

\subsection{Impact of Minimization on State Size and Comparison Efficiency}

\begin{table*}[t]
  \footnotesize
  \centering
  \begin{tabular}{| l ? l | r | r ? r | r | }
    \hline
    \multicolumn{6}{|l|}{\bf{DFA Subject Programs}} \\
    \hline
    \bf{State Type} & \bf{vs.} & \bf{ $\downarrow\Delta$ \% V} & \bf{$\downarrow\Delta$ \% E} & \bf{$\sim$ Inter, sec.} & \bf{$\sim$ Pred, sec.} \\
    \hline
    \acs{FS} & - & - & - & 4.03 & 265.91 \\
    \hline
    \acs{CC} & \acs{FS} & 70.37 & 29.47 & 1.41 & 4.09 \\
    \hline
    \acs{NN} & \acs{CC} & 0.02 & 0.01 & 1.41 & 4.04 \\
    \hline
    \acs{MN} & \acs{NN} & 0.10 & 0.05 & 1.35 & 4.05 \\
    \hline
    \multicolumn{6}{|l|}{\bf{EQBench Subject Programs}} \\
    \hline
    \acs{FS} & - & - & - & 0.79 & 5.56 \\[0pt]
    \hline
    \acs{CC} & \acs{FS} & 43.0 & 2.1 & 0.63 & 0.87 \\[0pt]
    \hline
    \acs{NN} & \acs{CC} & 0.0 & 0.0 & 0.58 & 0.9 \\[0pt]
    \hline
    \acs{MN} & \acs{NN} & 0.13 & 0.13 & 0.58 & 0.9 \\[0pt]
    \hline
  \end{tabular}
  \caption[Average percentage change between techniques]{Average percentage
    changes in $V$ and $E$ between each technique (columns 2--4), and average
    total runtime of state comparisons (columns 5,6).}%
  \label{tab:vars-preds-diff+z3-times}
\end{table*}

Table~\ref{tab:vars-preds-diff+z3-times} contains data for efficiency
evaluation, split over the two benchmark suites.  The table shows the average
percentage reduction in vertices and edges in Zones, comparing to the preceding
minimization algorithm (columns 2--4); and as a reduction in total runtime for
all comparisons between Zones and Interval states (column 5) and between Zones
and Predicates (column 6).

We aggregate the relative change in vertices and edges over all subject methods
for a more tractable comparison.  We use the percentage change to answer the
first part of \textbf{RQ1} related to state sizes.  The data show a large
reduction in the number of vertices between \acs{FS} and after applying
\acs{CC} algorithm.  The number of edges features a similarly significant,
though less dramatic reduction since they are compared without spurious
constraints.  On average, we see small reductions in the rest of the
comparisons.  The difference in vertex reductions versus edge reductions is due
to the reduced number of vertices required, contrasted with edge sparsity
arising from widening and merge operations which affects all representations.
However, as the small reduction of edges between \acs{MN} and \acs{NN} shows
that after removing more vertices from the subgraph, we remove more edges as
well.

The EQBench results mirror the reduction of edges and variables.  In the
EQBench benchmark suite, we see no reduction between \acs{CC} to \acs{NN}.
However, our final approach does remove vertices and edges from the previous
techniques of \acs{CC} and \acs{NN}.  This reduction is attributable to the
bisection enabled by our final technique which further reduces sparse graphs
based on the semantics of the changed constraint.

Addressing the second part of \textbf{RQ1}, we compare the average total
runtime of comparisons over the corpus of subject programs.  Columns 5 and 6 of
Table~\ref{tab:vars-preds-diff+z3-times} show the total runtime to execute all
domain comparisons, averaged over 5 executions, and broken down by benchmark
suite.  Between \acs{FS} and \acs{CC}, we see dramatic reductions in total
time.  As expected the remaining techniques show small improvements in
comparison time due to the minor reductions of vertices and edges shown in
columns 2--4.  The increase in time for Zone and Predicate comparison for the
EQBench subject programs is attributable to execution variance.  Overall, our
minimization algorithms reduce the size of Zone states and, in turn, improve
the efficiency of domain comparisons.

\subsection{Impact on Domain Comparison}

\begin{table*}[t]
  \footnotesize
  \centering
  \begin{tabular}{| c | r | r ? r | r | r | r |}
    \hline
    \multicolumn{7}{|l|}{\bf{DFA Subject Programs}} \\
    \hline
    \bf{State} & \bf{\(\succ\) Intervals} & \bf{\(=\) Intervals} & \bf{\(\succ\) Pred} & \bf{\(=\) Pred} & \bf{\(\prec\) Pred} & \bf{\(\prec\succ\) Pred}\\
    \hline
    \acs{FS} & 2898 & 1002 & 1464 & 237 & 167 & 2032\\[0pt]
    \hline
    \acs{CC} & 1194 & 2706 & 1324 & 1930 & 473 & 173\\[0pt]
    \hline
    \acs{NN} & 1191 & 2709 & 1322 & 1933 & 473 & 172\\[0pt]
    \hline
    \acs{MN} & 1164 & 2736 & 1305 & 1960 & 473 & 162\\[0pt]
    \hline
    \multicolumn{7}{|l|}{\bf{EQBench Subject Programs}} \\
    \hline
    \acs{FS} & 374 & 255 & 307 & 135 & 46 & 141\\[0pt]
    \hline
    \acs{CC} & 131 & 498 & 217 & 322 & 72 & 18\\[0pt]
    \hline
    \acs{NN} & 131 & 498 & 217 & 322 & 72 & 18\\[0pt]
    \hline
    \acs{MN} & 131 & 498 & 217 & 322 & 72 & 18\\[0pt]
    \hline
  \end{tabular}
  \caption[Summary of Domain Comparisons]{Summary of comparison between Zones
    and Intervals(2, 3), and between Zones and Predicates (4--7).}%
  \label{tab:precision-summary}
\end{table*}

\noindent{\bf{Comparable domains}} We compare Zones and Intervals invariants to
answer \textbf{RQ2}.  We break down the results by benchmark suites in
Table~\ref{tab:precision-summary}.  Columns 2 and 3 show the precision summary
of invariants between Zones and Intervals.

In the \acs{DFA} suite, using \acs{FS} to compare each domain, Zones compute
more precise invariants for approximately \(3/4\) of the total number of
invariants.  However, the ratio drops significantly to less than a third,
(\(31\%\)), when using the \acs{CC} technique.  Our final technique \acs{MN}
lowers the percentage of invariants where Zones are more precise to about
\(30\%\) of all computed invariants.  Furthermore, our techniques demonstrate
the preponderance of invariants where Zones and Intervals are equivalent.  We
see similar results when considering the methods of the EQBench suite.  Using
the full state to compare Zones and Intervals, we see Zones compute a majority
of more precise invariants, about \(59\%\).  However, using any one of our
minimization techniques moves the proportion of more precise invariants to
\(21\%\).  We attribute the lack of further reduction with later techniques to
the preponderance of non-integer operations in the EQBench suite.

\noindent{\bf{Incomparable Domains}} Additionally, with respect to
\textbf{RQ2}, we compare Zones to Predicates to evaluate whether our techniques
minimize the number of incomparable invariants computed between the two
domains.  Since Zones and Predicates are incomparable, we
consider all comparison categories denoted here as: \(\succ Pred\) for Zones more
precise than Predicates; \(= Pred\) for Zones equivalent to Predicates;
\(\prec Pred\) for Zones less precise than Predicates; and \(\prec\succ Pred\) for Zones
and Predicates being incomparable.

Columns 4--6 of Table~\ref{tab:precision-summary} summarizes the distribution of
relative precision for Zones and Predicates over the computed invariants of the
subject programs.  Unlike Zones, Predicates cannot represent arbitrary integer
constants; Predicates are limited to the \textit{a priori} chosen predicate
elements.  However, Predicates can represent disjoint ranges of values which
Zones, and other numerical domains, cannot.  As such, when using \acs{FS}, we
see a high percentage of invariants fall into either \(\succ Pred\) and
\(\prec\succ Pred\).  The \(\succ Pred\) makes up about \(38\%\) of invariants.  Similarly,
\(\prec\succ Pred\) weighs in at \(\sim52\%\) of invariants.

When, applying \acs{CC}, the percentage of incomparable invariants drops
significantly to \(4\%\).  \(\prec Pred\) comprise \(12\%\) of invariants, up from
\(4\%\).  Similarly, the percentage of \(\succ Pred\) drops from \(38\%\) to
\(34\%\).  Finally, equality between the two domains significantly increased
from \(6\%\) to about \(49\%\).  These trends continue for each technique.
Each technique shifts the distributions of invariants from \(\succ Pred\) and
\(\prec\succ Pred\) to \(= Pred\) and \(\prec Pred\).  In \acs{MN} state, Zones are more
precise for about \(33\%\) of invariants, down from \(38\%\); Zones are equal
to Predicates for about \(50\%\) of invariants, up from \(6\%\); and Zones and
Predicates are incomparable for about \(4\%\) of invariants, down from
\(52\%\).  For each technique, Zones less precise than Predicates remained at
\(12\%\), up from \(4\%\) compared to \acs{FS}, between the techniques.
Considering the EQBench methods, we observe similar results for Zones and
Predicates.  Using the \acs{CC} technique, we see a significant shift in the
distribution of invariants.  However, we do not see any further distribution
shifts in this program set.  We attribute this to the fact that the EQBench
methods consist of many non-integer operations.




Clearly, our techniques reduce the percentage of incomparable invariants,
enabling, for example, more accurate comparison between Zones and incomparable
domains, such as Predicates.  While not the goal of this study, the comparable
results confer merit to previous research which anecdotally mentions: the
majority of computed invariants are interval
valued~\cite{gange-2021-fresh-look,howe-2009-logah}.  This improved accuracy
would be especially valuable in adaptive analysis approaches where a heuristic
decides which abstract domain to utilize for a specific block of code.


\subsection{Threats to Validity}

\noindent{\bf{External Threats}} The subject programs from previous
research~\cite{badihi-2021-eqbench,sherman-2015-exploit-domain} were extracted
from real, open-source projects, each with a high number of integer operations.
The EQBench suite consists of predominately numerical programs but demonstrate
the generalizability of our results.  Other concerns include the choice of
Predicate elements and lack of narrowing which could influence the precision
counts between Zones and Predicates and between Zones and Intervals,
respectively.  However, since we examine only the trend of the different
categories, the exact precision does not affect our conclusions.

\noindent{\bf{Internal Threats}} To mitigate internal threats to validity, we
developed a large test suite, \(703\) unit tests, to ensure our implementation
is correct.  The test suite contains numerous tests which check the consistency
of the partial order over Zones and Intervals.  Furthermore, we developed and
manually verified tests to check comparison between Zones and Predicates.
Specifically, the test suite contains manually verified tests which use real
subject programs to test the correctness of the analyses and their comparisons.



\section{Related Work}\label{sec:related}%

We have mentioned our spurious reduction technique is based on the work by
Larsen \textit{et~al.}~\cite{larsen-1997-effic-verif}.  Their algorithm removes
all redundant constraints without removing variables, reducing the overall
number of \acl{LII} but it does not reduce the number of variables.  Along
similar lines, Giacobazzi \textit{et~al.}~\cite{giacobazzi-2002-domain-compr}
proposed techniques for abstract domain compression for complete finite
abstract domains.  That is, it reduces the number of constraints within the
logical formula without altering the approximation of the abstract domain.  Our
techniques extract subsets of the state for specific verification tasks.


Our approach, \acf{CC}, resembles the slicing technique of the Green solver
interface~\cite{visser-2012-green} and split normal form introduced by Gange
\textit{et~al.}~\cite{gange-2021-fresh-look,gange-2016-exploit-spars}.  Our
approach differs from Green in application and restriction to Zone constraints.
Slicing can extract connected constraints by what variables are present in the
set of constraints.  However, we can exclude transitive relations between
variables because within Zones, not all variables are modified by the
introduction of a new constraint.


We base our methodology on previous work on new abstract domains which provide
a comparison of the new domain against other known similar or comparable
domains.  These comparisons can be categorized into two predominate strategies.
The first, domains are compared via a known set of properties over benchmark
programs~\cite{gange-2021-fresh-look,gurfinkel-2010-boxes,howe-2009-logah,laviron-2008-subpol,logozzo-2010-pentag}.
The second, domains are compared via logical entailment of the invariants
computed at program
points~\cite{mine-2004-weakl-relat,sherman-2015-exploit-domain}.  In the first
case, the comparison is straightforward.  In the latter case, as this work
demonstrates, the precision between two domains can depend on the set of
invariants used to perform the comparison.





\section{Conclusion and Future Work}\label{sec:concl}%

We proposed several techniques which identify a minimal set of changed
inequalities for the Zones abstract domain.  Our techniques improve upon
existing techniques such as Green's slicing~\cite{visser-2012-green} technique
which further reduces the number of dependent variables within a changed set of
inequalities.  We empirically evaluated our techniques and showed improvements
of efficacy and efficiency.  Concretely, the changed subgraph of Zones is
equivalent to Intervals in more than \(70\%\) of computed invariants, a result
commented on but never demonstrated in previous
research~\cite{gange-2021-fresh-look,howe-2009-logah}.  Moreover, our
techniques significantly reduced the incomparable invariants found when
comparing two incomparable domains, resulting in a clearer picture of the
relative precision between the two domains.  Furthermore, the reduction in
variables improved the efficiency of domain comparison, reducing average total
runtime of incomparable domain comparisons by \(98\%\).  While evaluated within
the context of \acs{DFA} frameworks, we presented general algorithms which, we
believe, are applicable in other areas of formal methods such as model checking
and symbolic execution.

\noindent{\bf{Future Work}} We intend to extend this work to include additional
relational domains.  Specifically, enabling comparison between two relational
domains, such as Zones and Octagons, is an interesting avenue to pursue.  Since
the techniques improve accuracy in comparison between domains they could be
beneficial for adaptive static analysis techniques which selectively use the
best abstraction.  We believe this work also opens up possibilities of
comprehensive studies which empirically validate several abstract domains and
their partial ordering.  Specifically, it would be interesting to see
comprehensive comparisons between Predicate domains and Zones.



\section*{Acknowledgments}

The work reported here was supported by the U.S.  National Science Foundation
under award CCF-19-42044.


\nocite{tange_2022_7465517}
\bibliographystyle{splncs04}
\bibliography{bibfile}

\end{document}